\documentclass[aps,twocolumn]{revtex4}
\usepackage{epsfig}
\usepackage{graphicx}
\usepackage{amsmath,amssymb,amsfonts,slashed}
\usepackage{array}
\usepackage{url}

\usepackage[colorlinks=true, linkcolor=blue, citecolor=blue, urlcolor=blue]{hyperref} 
\usepackage[utf8]{inputenc}
\usepackage{bm}
\usepackage{multirow}
\usepackage{float}
\usepackage{lineno}
\usepackage{xspace}
\usepackage{ulem}
\usepackage{nameref}
\usepackage[usenames,dvipsnames]{color}

\newcommand{\bef}{\begin{figure}}
\newcommand{\eef}{\end{figure}}
\newcommand{\bc}{\begin{center}}
\newcommand{\ec}{\end{center}}

\newcommand{\be}{\begin{equation}}
\newcommand{\ee}{\end{equation}}
\newcommand{\bea}{\begin{eqnarray}}
\newcommand{\eea}{\end{eqnarray}}

\def\ba{\begin{eqnarray}}
\def\ea{\end{eqnarray}}
\makeatletter
\renewcommand\thefootnote{\textcolor{blue}{\@fnsymbol\c@footnote}}
\makeatother
\begin{document}
\title{\textbf{Relative Hadron Yields in HRG With Medium Modification}}
 
\author{Nasir Ahmad Rather\footnote{\texttt{\href{mailto:nasir@example.com}{nasirrather345@gmail.com}}}, Sameer Ahmad Mir, Iqbal Mohi Ud Din, Saeed Uddin\footnote{Corresponding author: \texttt{\href{mailto:saeed@example.com}{suddin@jmi.ac.in}}}} 
\affiliation{Department of Physics, Jamia Millia Islamia, New Delhi, India}
\begin{abstract}
In the framework of a constituent quark mass model,  the modified baryon masses are incorporated into the hadron resonance gas (HRG) based analysis of the like mass particle ratios in ultra relativistic nucleus-nucleus collisions (URNNC) over a wide range of collision energy. In addition we have incorporated an essential feature of the hadronic interaction at short distance, i.e. the hard-core repulsion by using the standard excluded volume type approach. We have extracted the chemical freeze-out conditions. The resulting freeze-out line in our case is compared with those obtained earlier using different model approaches. The correlation between $k^{-}/k^{+}$ and $\bar p/p$ ratios is also studied.

\end{abstract}
\maketitle
\section{Introduction}
\label{intro}
Properties of strongly interacting matter  have been a subject of great interest   for over several decades as their study may reveal some deep insight into the nature of the evolution of matter during early universe when a transition may have occurred inside such a matter involving two distinctly different phases. This transition between a hot and dense deconfined phase of quarks and gluons to a system of hadrons, where quarks and gluons are confined, also called hadron resonance gas (HRG), can occur during the expansion and cooling of such a matter. At sufficiently high temperature the deconfined matter may also reach a chirally symmetric phase and subsequent to its expansion and cooling, it may transit into a dynamically chiral symmetry breaking phase followed by hadronization stage. The temperature (T) and chemical potential $(\mu_B)$, are often the two thermodynamic parameters  that are required to drive such phase transitions. Hence exploring the quantam chromodynamics (QCD) phase diagram  is one of the main goals of URNNC experiments. The QCD phase diagram is  usually depicted in the T - $\mu_{B}$ plane which shows a possible transition from extreme energy density and temperature phase dominated by partonic degrees of freedom  to a phase where the relevant degrees of freedom are hadronic \cite{muzinger1,muzinger2,philipsen,ratti,adams,qcd:lattice,wilchalk}.

This transition from a state of quasi free light quarks to a state of mesons and baryons in the early universe is expected to have  occured at $T_{c}$ $\sim$ $100-200 $ MeV \cite{witten:1984}. The lattice QCD calculations predicts this phase transition to occur at a temperature $\sim 140-170 $ MeV \cite{bernad,lee}. The calculations from lattice QCD (LQCD) also indicate a rapid cross over to quark-hadron phase for a system produced at $\mu_{B} \approx 0 $ while for systems produced at finite $\mu_{B}$, a phase transition of first order may occur \cite{nature,ejiri}. To produce such a partonic state of matter called quark-gluon plasma (QGP), a number of experiments have been performed by colliding large sized  nuclei at ultra-relativistic energies \cite{arsene,alver}. Typically the thermal temperature of the  hot and dense matter produced in these experiments extracted in the framework of thermal models falls in the range $100-200$ MeV \cite{saeed}. In the limit of chiral symmetry the expectation value of the quark condensate $<\bar \psi \psi >$ vanishes and opposite parity states (chiral partners) are degenerate. However, it is also expected that due to the interactions, the masses of the constituent quarks in hadrons will be significantly affected resulting in the lowering of the  baryon masses after hadronization.
The lattice simulations however suggest that the
confinement-deconfinement and chiral phase transitions may  occur at the same temperature T $\sim$ 160 MeV \cite{arsene1,meyer}. At this temperature (T) only the three lightest quark flavours, the up ($\sim$ 5.1 MeV), down ($\sim$ 9.1 MeV) and strange ($\sim$ 135.7 MeV), influence the thermodynamic properties of the system. The ideal or non interacting HRG
model is somewhat successful in reproducing the zero chemical
potential LQCD data of bulk properties of the QCD matter
at low temperatures, $T<150$ MeV \cite{bazavov,fodor}. However,
disagreement between LQCD data and ideal HRG model calculations have been observed at higher temperatures. Considering excluded volume correction, which mimics repulsive
interaction in HRG model, one can improve the picture in the
crossover temperature region, $T \sim 140-190$ MeV \cite{das}. The need to incorporate short-range repulsive interactions arises from fact that in the hot and dense systems created in the URNNC (such as at AGS, SPS, RHIC and LHC energies) the particle number densities are very high and the repulsive hard-core interaction can significantly affect the multiplicities of the hadrons produced after the final break-up of the system, also called freeze-out.
The particle number densities calculated in the framework of the thermal models gets suppressed by incorporating repulsive interactions into the HRG model via excluded volume type correction. In many Excluded Volume HRG (EVHRG) models, effects of Van der Waals type hadronic repulsions at short distances are considered but long distance attractive interactions are ignored \cite{samanta}. In the present work to account for chiral symmetry which is a feature of QCD langragian we have used the SU(3) Nambu Jona-Lasinio (NJL) model. By employing this model to describe baryonic matter, one can effectively incorporate three key aspects of QCD: chiral symmetry, mass generation through explicit and spontaneous symmetry breaking and the binding of ground state baryonic matter, resulting from the changes in the condensate. In NJL model it is possible to construct a comprehensive physical picture from
the calculations performed  both at $T=0$ MeV and at finite
temperatures and densities \cite{hatsuda,t.j}. Moreover in NJL model,
hadrons can be understood as dynamically generated states
from multi-quark re-scattering, thus offering  a non-perturbative mechanism for an effective confinement \cite{torres}.

The masses of constituent quarks \textit{(u,d,s)} are computed by using  SU(3) NJL model and then incorporated into constituent quark mass model to obtain T and $\mu$ dependent baryon masses. The masses of mesons have been taken to be almost constant since the Goldstone modes show a weak dependence on temperature as they are associated with SU(3) chiral symmetry breaking and moreover this  approach is in agreement with the explicit model calculations incorporating chiral symmetry \cite{hatsuda,kadam,lutz}. Since baryon's constituent quark masses depend on temperature (T) and chemical potential ($\mu$), thus the  baryon masses that appear in the partition function of EVHRG model for computing the particle number density, also become T and $\mu$ dependent. The EVHRG based models have been generally successful in describing the ratios of hadron yields from low SIS and AGS energies all the way upto SPS, RHIC and LHC  energies \cite{hatsuda1,muzinger,rafelski}.

Earlier it has been shown that by using constant   mass for point-size hadrons and only two thermal parameters (T, $\mu_{B}$), a reasonable description of particle ratios measured in such collisions may be obtained \cite{sahoo,bhatta2020}. However, in this paper our focus is on calculating  ratios of like  mass baryons and mesons  over a wide range of collision energies in the URNNC (from AGS to LHC), by incorporating T and $\mu$ dependent baryon masses in the framework of EVHRG model and extract the freeze-out conditions of the hot and dense HRG formed in such collisions. The freeze-out line obtained through our calculation is compared with some previous cases.  

We organize the paper as follows. In Sec. II we briefly describe the NJL model (providing the 
T, $\mu$ dependent baryon masses) and thermodynamically consistent EVHRG model. In Sec. III we present the result and discussions on particle ratios and the freeze-out conditions. Finally we summarize and conclude in Sec. IV.

\section{ MODEL DESCRIPTION}
NJL model was orginally proposed as an attempt to explain the nucleon mass by spontaneous breaking of chiral symmetry \cite{nambu}.
In present form it  is an effective model of QCD which respects all the symmetries of QCD and describes the low energy interaction of fermions by contact vertex \cite{torres}. Since in low energy region the number of active flavors is three, so the effective langragian of QCD contains three lower mass quarks. The interaction among these flavors is constrained by $SU _{L} (3)\bigotimes SU_{R}(3)$ chiral symmetry, explicit symmetry breaking due to current quark masses and the $U_{A}(1)$ breaking due to axial anomaly. A standard form of SU(3) NJL model as such reads as \cite{hatsuda,g. ala,kalevensky}:
\begin{align}
\mathcal{L}_{\text{NJL}} &= \bar{q}(i\slashed{\partial} - \hat{m})q 
+ \frac{g_S}{2} \sum_{a=0}^{8} \left[ (\bar{q}\lambda_a q)^2 + (\bar{q}i\gamma_5\lambda_a q)^2 \right] \notag \\
&\quad - g_D \left[\det \bar{q}(1 + \gamma_5)q + \det \bar{q}(1 - \gamma_5)q \right]
\label{Eqn. 1}
\end{align}

Here $q=(u,d,s)$ is the quark triplet, $\hat{m}$ = ($m_{0u}$, $m_{0d}$, $m_{0s}$) is the current quark mass matrix which explicitly breaks SU(3)
chiral symmetry. The $\lambda^{a}$
 are Gell-Mann matrices. The determinant term is chiral invariant but breaks $U_{A}(1)$ symmetry and is reflection of axial annomaly in QCD.

In the mean field (Hartee-Fork) approximation, the  grand thermodynamical potential density can be obtained from Eqn.~\eqref{Eqn. 1} which leads to three  gap equations for the constituent quark mass of the following type:

\begin{equation}
m_u^* = m_{0u} - 4G_s \langle \overline{\psi}_{u}\psi_{u} \rangle - 2K_{s} \langle \overline{\psi}_{d} \psi_{d} \rangle \langle \overline{\psi}_{s} \psi_{s}  \rangle
\end{equation}

\begin{equation}
m_d^* = m_{0d} - 4G_s \langle \overline{\psi}_{d}\psi_{d} \rangle - 2K_{s} \langle \overline{\psi}_ {s} \psi_ {s} \rangle \langle \overline{\psi}_ {u} \psi_ {u} \rangle
\end{equation}

\begin{equation}
m_s^* = m_{0s} - 4G_s \langle \overline{\psi}_{s}\psi_{s} \rangle - 2K_{s} \langle \overline{\psi}_{u} \psi_{u}\rangle \langle \overline{\psi}_{d} \psi_{d}\rangle
\end{equation}
here $(\overline{\psi} \psi)=\alpha_{q}$ (where $q=u,d,s$) is the quark condensate and is given as:
\begin{equation}
 \alpha_{q}=-\frac{2N_{c}}{2\pi^{3}}\int_{0}^{\Lambda}\frac{m^{*}_{q}d^{3}p}{E_{q}}[1-f_{q}(T,\mu)-\overline{f}_{q}(T,\mu)] 
 \label{Eqn. 5}
\end{equation}
$E_{q} = \sqrt{p^{2}+m_{q}^{*^{2}}}$ is energy of quasi-particle and $f_{q}(T,\mu)$ is Fermi-Dirac distribution of $q^{th}$ quark flavour.
\begin{table}
\renewcommand{\arraystretch}{1.5}
\begin{tabular}{|c|c|}

\hline

Parameters & Values \\
\hline
$m_u=m_d$ & $5.1$ MeV \\[1pt]

$m_{s} $& $135.7$ Mev\\
$\Lambda$ & $631.4$ MeV \\

$G_{s}\Lambda^{2}$ & $3.67$ MeV \\

$K_{s}\Lambda^{5}$ &$-9.31$ Mev\\
\hline
\end{tabular}
\caption{Parameters for  SU(3) NJL model.}
\label{TABLE 1}
\end{table}
Here $\Lambda$ is the momentum cut-off, $G_{s}$ and ${K_{s}}$  are coupling constants, $m^{*}_{u}, m^{*}_{d}, m^{*}_{s}$ represents effective constituent quark masses of three light flavours, which are T and $\mu$ dependent while $m_{0u},m_{0d},m_{0s} $ denote their current quark masses, respectively. The $N_{c}=3$ is the number of quark colours. The values of the parameters used in our calculations are shown in  Table I. The parameters are chosen such that the  calculated baryon masses for T $\sim 0$ agree well with the known vacuum  baryon mass spectrum. The parameters are taken from \cite{hatsuda,alberico}.
Using the gap equations \cite{hatsuda}, the  masses of baryons as a function of temperature (T) chemical potential ($\mu$) are given in Table II.
\begin{table}[htbp]
    \centering
\renewcommand{\arraystretch}{1.5}
    \label{tab:baryon_mass_formulas}
    \begin{tabular}{|p{1.8cm}|p{6.6cm}|}
        \hline
        \textbf{Baryons} & 
    \textbf{$\mathbf{M}_{\mathbf{B}}$} \\
         \hline
        {p} & $M_{0} + 2m_u^* + m_d^* + \frac{a}{2} \left( \frac{2}{m_u^*} + \frac{1}{m_d^*} \right) + 4b \left( \frac{1}{4m_u^*2} - \frac{1}{m_u^* m_d^*} \right)$ \\[6pt]
       $\Delta^{+}$ & $M_{0} + 2m_u^* + m_d^* + \frac{a}{2} \left( \frac{2}{m_u^*} + \frac{1}{m_d^*} \right) + 4b \left( \frac{1}{4m_u^*2} + \frac{1}{2m_u^* m_d^*} \right)$ \\[6pt]
        $\Delta^{++}$ & $M_{0} + 3m_u^* + \frac{3a}{2} \cdot \frac{1}{m_u^*} + 3b \cdot \frac{1}{m_u^*2}$ \\[6pt]
        $\Sigma^{+}$ & $M_{0} + 2m_u^* + m_s^* + \frac{a}{2} \left( \frac{2}{m_u^*} + \frac{1}{m_s^*} \right) + 4b \left( \frac{1}{4m_u^*2} - \frac{1}{m_u^* m_s^*} \right)$ \\[6pt]
       $\Sigma^{*+}$ & $M_{0} + 2m_u^* + m_s^* + \frac{a}{2} \left( \frac{2}{m_u^*} + \frac{1}{m_s^*} \right) + 4b \left( \frac{1}{4m_u^*2} + \frac{1}{2m_u^* m_s^*} \right)$ \\[6pt]
        $\Xi^{0}$ & $M_{0} + m_u^* +m_d^* +m_s^* + \frac{a}{2} \left( \frac{1}{m_u^*} + \frac{2}{m_s^*} \right) + 4b \left( \frac{1}{4m_s^*2} - \frac{1}{m_{v}m_s^*} \right)$ \\[6pt]
        $\Xi^{*+}$ & $M_{0} + m_u^* + 2m_s^* + \frac{a}{2} \left( \frac{1}{m_u^*} + \frac{2}{m_s^*} \right) + 4b \left( \frac{1}{4m_s^*2} + \frac{1}{2m_u^*m_s^*} \right)$ \\[6pt]
        $\Lambda^0$ & $M_{0} + m_u^* + m_d^* + m_s^* + \frac{a}{2} \left( \frac{1}{M_u^*} + \frac{1}{M_d^*} + \frac{1}{M_s^*} \right) - 3b \cdot \frac{1}{m_u^*m_d^*}$ \\[6pt]
        $\Omega^-$ & $M_{0} + 3m_s^* +\frac{3a}{2} \cdot \frac{1}{m_s^*} +3b \cdot \frac{1}{m_s^*2} $ \\[3pt]
        \hline
    \end{tabular}
    
    \caption{Mass formula's for various baryons in the constituent quark model. Here the value of $M_{0}$= -56.4 MeV, $a= (175.2$ $MeV)^2$ and $ b= (176.4$ $MeV)^3$. For elaboration see \cite{hatsuda}}
\end{table}

The  baryon masses thus obtained are used in our statistical thermal model based description of the hot and dense hadronic system formed in URNNC under the assumption that the hadronic species achieve a reasonably high degree of thermal and chemical equilibrium \cite{parra}. To incorporate the hard-core baryonic repulsion in the HRG model, the approach of Rischke et. al. \cite{rischke} is used. This approach is in the strict sense thermodynamically consistent. Due to sufficiently large and non-conserved number of hadrons produced in URNNC we can effectively employ the grand canonical formalism. Moreover, it has a advantage over canonical ensemble in explaining the complete thermodynamic description of such systems \cite{v3}. The expression for the grand canonical partition function for the ideal  point-like hadronic specie can be written as:
\begin{equation}
lnZ^{id}(T,\mu,V)=\frac{Vg}{2\pi^{2}}\int_{0}^{\infty}p^{2}       ln\left[1 \pm e^{-(E-\mu)/T}\right]dp
  \label{Eqn. 6}
\end{equation}

The + sign corresponds to fermions while the - sign to the bosons. The  $E$ represents the energy of the  point-like hardonic specie and is given by
\[
E = \sqrt{p^2 + m^{2}}.
\]
Here, \textit{m} is the mass of the given hadron which  in case of baryon (anti-baryon) is the \textit{T}, \textit{$\mu$} dependent, while $g$, \textit{T, V}  denotes  spin-isospin degeneracy, temperature and physical volume of the system, respectively. Further $\mu$ denotes the chemical potential for the  hadronic specie and is  given by:
\begin{equation}
 \mu=B\mu_{B}+S\mu_{S}+Q\mu_{Q} 
 \label{Eqn. 7}
\end{equation}
Here, $B$, $S$, and $Q$ are the baryon number, strangeness, and electric charge, whereas $\mu_B$, $\mu_S$, and $\mu_Q$ represent the baryon chemical potential, strangeness chemical potential and electric chemical potential, of a given hadronic specie, respectively.

Using Eqn.~\eqref{Eqn. 6} one can get the pressure for the case of HRG phase as:

\begin{equation}
\begin{array}{rcl}
P^{id}(T,\mu) &=& \lim_{V\to\infty} T \frac{\ln Z^{id}(T,\mu,V)}{V} \\ \\
&=& \frac{Tg}{2\pi^{2}}\int_{0}^{\infty}p^{2} ln\left[1\pm e^{-(E-\mu)/T}\right]dp\\ 
\end{array}
\label{Eqn. 8}
\end{equation}

The number density of the point-like particles can be obtained in thermodynamically consistent manner as:

\begin{equation}    
n^{id}(T,\mu) = \left(\frac{\partial P^{id}}{\partial \mu }\right)_T
\label{Eqn. 9}
\end{equation}
In terms of the corresponding canonical partition function one can also write the ideal grand canonical partition function as:
\begin{equation}
   Z_{GC}^{id}(T,\mu,V)=\sum_{N=0}^{\infty}e^{\frac{\mu N}{T}}\mathcal{Z}_{C}^{\textit {id}} (T,\mu,N)
   \label{Eqn. 10}
\end{equation}
Here $\mathcal{Z}_{C}^{id} (T,\mu ,N )$ represents the ideal canonical partition function for N number of particles.  Incorporating the excluded volume effect, the modified grand canonical partition function can be written in following form \cite{rischke}:
\begin{equation}
 Z_{GC}^{excl}(T,\mu,V)=\sum_{N=0}^{\infty}e^{\frac{\mu N}{T}}\mathcal{Z}_{C}^{\textit {id}}(T,N,V-bN) 
 \label{Eqn. 11}
\end{equation}
The quantity $bN$ (where $b=\frac{16 \pi r^3}{3}$)
represents the excluded volume 
arising due to hard-core repulsive interaction among the baryons and $V-bN$
denotes the available volume
in the system. By taking the Laplace transform of Eqn.~\eqref{Eqn. 11} one can obtain the pressure of the system: 
\begin{equation}
\hat{Z}_{GC}^{excl}(T, \mu, \zeta)=\int_{0}^{\infty} e^{-\zeta V}   Z_{GC}^{excl}(T,\mu,V)dV
\label{Eqn. 12}
\end{equation}
for the finiteness of the integral we must have:
\begin{equation}
    \zeta=  \lim_{V\to\infty} \frac{\\ln Z_{GC}^{\textit{excl}}(T,\mu,V)}{V}
    \label{Eqn. 13}
\end{equation}
which gives $T\zeta=p^{excl}(T,\mu)$.
Using Eqn.~\eqref{Eqn. 11} in Eqn.~\eqref{Eqn. 12} we get:
\begin{equation}
\hat{Z}_{GC}^{excl}(T,\mu,\zeta)=\int_{0}^{\infty} e^{-\zeta V} dV\sum_{N=0}^\infty e^{\frac{\mu N}{T}} \mathcal{Z}_{C}^{excl}(T,\mu,V-bN)
\label{Eqn. 14}
  \end{equation}
  Defining a new variable $\mu^*$ (as the effective chemical potential) and interchanging the order of summation and integration one can show that \cite{rischke,sam}:
  \begin{equation}
   Z_{GC}^{excl}(T,\mu,V)=\mathcal{Z}_{GC}^{id}(T,\mu^*,V)   
   \label{Eqn. 15}
  \end{equation}
  where 
  \begin{equation}
\mu^*=\mu-b\zeta T=\mu-bp^{excl}(T,\mu)
\label{Eqn. 16}
  \end{equation} 
From Eqn.~\eqref{Eqn. 15} we also get
  \begin{equation}
    \lim_{V\to\infty} \frac{T}{V}{\ln Z_{GC}^{excl}(T,\mu,V)}= \lim_{V\to\infty} \frac{T}{V}{\ln Z_{GC}^{id}(T,\mu^*,V)} 
    \label{Eqn. 17}
  \end{equation}
 \hspace{1.3cm}    $ \Rightarrow p^{excl}(T,\mu,V)=p^{id}(T,\mu^{*},V)$ 
\vspace{0.5cm} \\
  By taking the derivative of the pressure we can obtain the number density of finite size particle in a thermodynamically consistent manner as follows: 
\begin{equation}
  n^{excl}(T,\mu)=\left (\frac{\partial p^{excl}(T,\mu)}{\partial \mu}\right)_T =\frac{n^{id}(T,\mu^{*})}{1+bn^{id}(T,\mu^{*})}
  \label{Eqn. 18}
\end{equation}
For a multi-component system we can generalize the above equation in order to get the number density of the $j^{\text{th}}$ hadronic specie as:
\begin{equation}
 n_{j}^{excl}(T,\mu_{j})=\frac{n_{j}^{id}(T,\mu^{*}_{j})}{1+\sum_{i} b_{i}n_{i}^{id}(T,\mu_{i}^{*})}   
\end{equation}
The decay contribution of the heavier resonances upto omega ($\Omega$) mass has also been taken into account \cite{vovchenko}. The total particle yield is thus the sum of the primordial  (thermal) yield and decay contributions after the chemical freeze-out as given below:
\begin{equation}
 n_{j}^{tot}=n_{j}^{th}+n^{decay} =n_{j}^{th}+\sum_{i }n_{i}^{th} Br(i \rightarrow j )  
\label{Eqn. 19}
\end{equation}

\noindent where $ n_{j}^{th} (n_{i}^{th})$ is the primordial particle number density which is calculated using Eqn.~\eqref{Eqn. 18}, while $ Br(i\rightarrow j)$ is the branching ratio of particle $\textit{i}$ decaying into particle $\textit{j}$ \cite{hagiwara}.
 \section{Results and Discussion} 
We will consider baryon-antibaryon, meson-meson and meson-baryon interaction as only attractive in nature. The hard-core repulsive interaction is assumed to exist for baryon-baryon and antibaryon-antibaryon pairs. For simplicity we have used the Boltzmann distribution function for all hadrons in the analysis.The strange chemical potential in Eqn.~\eqref{Eqn. 7}, is fixed for the given values of T and $\mu_B$ by applying the  strangeness conservation criteria, while the electric chemical potential $\mu_Q$ is fixed by the charge to baryon ratio of the colliding system, which turn out to be almost same $\sim$ 0.4, over the wide range of the collision energies.
We are thus left with only one independent chemical potential i.e. the baryon chemical potential, $\mu_{B}$. 

The chemical potential of the anti-particles are defined as the negative of their corresponding particles chemical potential. The freeze-out values of temperature and baryon chemical potential can be fixed by using the following parameterization introduced by Cleyman and also suggested by a number of studies   \cite{cleyman,koch,cleymans freezeout paper,cp}:
\begin{equation}
   T(\mu_{B})=c-d\mu_{B}^2-e\mu_{B}^4
   \label{Eqn.21}
\end{equation}
\vspace{-8mm}
\begin{equation}
  \mu_{B}(\sqrt{s_{NN}})=\frac{f}{1+g\sqrt{s_{NN}}} 
  \label{Eqn.22}
\end{equation}
 \vspace{-3mm}

The above ansatz have been widely used to study the properties of hot and dense matter formed in URNNC  at different centre of mass energies in various models, but with different values of the ansatz parameters \cite{letessier,biswas}. The values of the ansatz parameters in our case are fixed by obtaining the best theoretical fit to the experimental data. For this purpose we have used the antiproton to proton ratio data. The choice of this data set is guided by the fact that proton (antiprotons) are the most abundantly produced baryons (antibaryons) in the system. Along with the parameters in  Eqn.~\eqref{Eqn.21} and Eqn.~\eqref{Eqn.22}, we have treated the hard-core baryonic radius also as a free parameter in order to obtain the best fit to the experimental data over a wide range of RHIC-STAR and ALICE-LHC energies. To calculate the other like mass anti-baryon to baryon ratios and relative mesonic yields we have used the same set of parameters to test if they can simultaneously (and at least reasonably) describe these ratios as well \cite{cleyman}. For the goodness of fit we have used the following minimum chi squared test:
\begin{equation}
  \chi^{2}= \sum_{i=1}^{N}\frac{(R_{j}^{th}-R_{j}^{exp})^{2}}{\sigma_{j}^{2}} 
\end{equation}
Here $R_{j}^{th}$, $R_{j}^{exp}$ and $\sigma_{j}$ denotes theoretical value,  experimental value and error, respectively for the given $\sqrt {s_{NN}}$. The values of constants in above ansatzs of temperature {T} and baryon chemical potential $(\mu_{B})$ turn out to be:

\vspace{0.5mm}

   $c = 145 \pm 1.3\ \text{MeV},\ d = 0.17\pm 0.01\ \text{MeV}^{-1}$,  
   
   \vspace{0.15cm}
   
   $e = 0.015\pm 0.08 \ \text{MeV}^{-3}, \ f = 1180\pm 16.0 \ \text{MeV}$,\
      \vspace{0.15cm}

   $g= 0.28\pm 0.01 \ \text{GeV}^{-1} $

\begin{figure}[H]
    \centering
\includegraphics[width=0.45\textwidth, height=0.2\textheight]{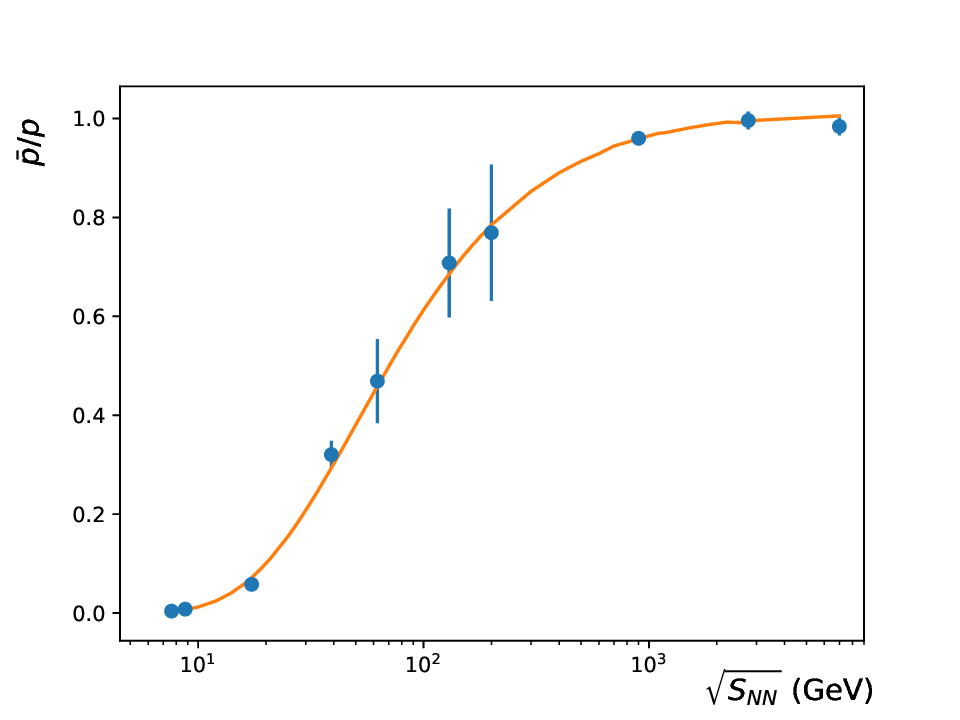}
    \caption{$\bar{p}/p$ dependence on $\sqrt{s_{NN}}$. Experimental data is taken from \cite{63,64,65,66,67,68}.} \label{fig:1}
\end{figure}

\begin{figure}[H]
    \centering
\includegraphics[width=0.45\textwidth, height=0.2\textheight]{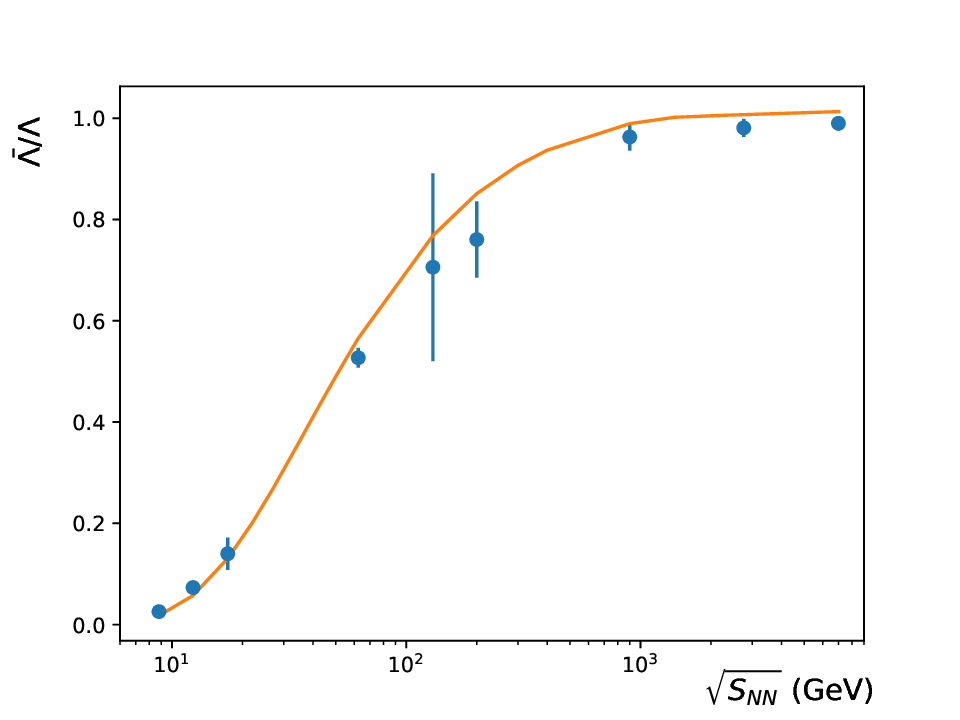}
    \caption{$\bar{\Lambda}/\Lambda$ dependence on $\sqrt{s_{NN}}$. Experimental data is taken from \cite{65,67,69,70,71}.}
   \label{fig:2}
\end{figure}

\begin{figure}[H]
    \centering
\includegraphics[width=0.45\textwidth, height=0.2\textheight]{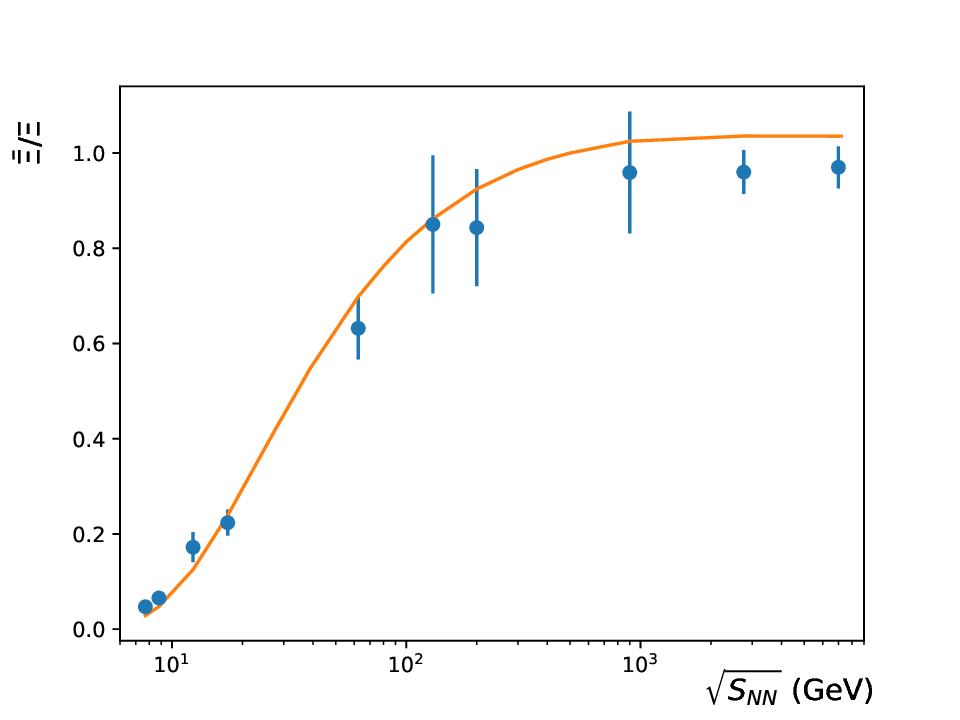}
    \caption{$\bar{\Xi}/\Xi$ dependence on $\sqrt{s_{NN}}$. Experimental data is taken from \cite{65,67,69,70,71}}
   \label{fig:3}
\end{figure}

\begin{figure}[H]
    \centering
\includegraphics[width=0.45\textwidth, height=0.2\textheight]{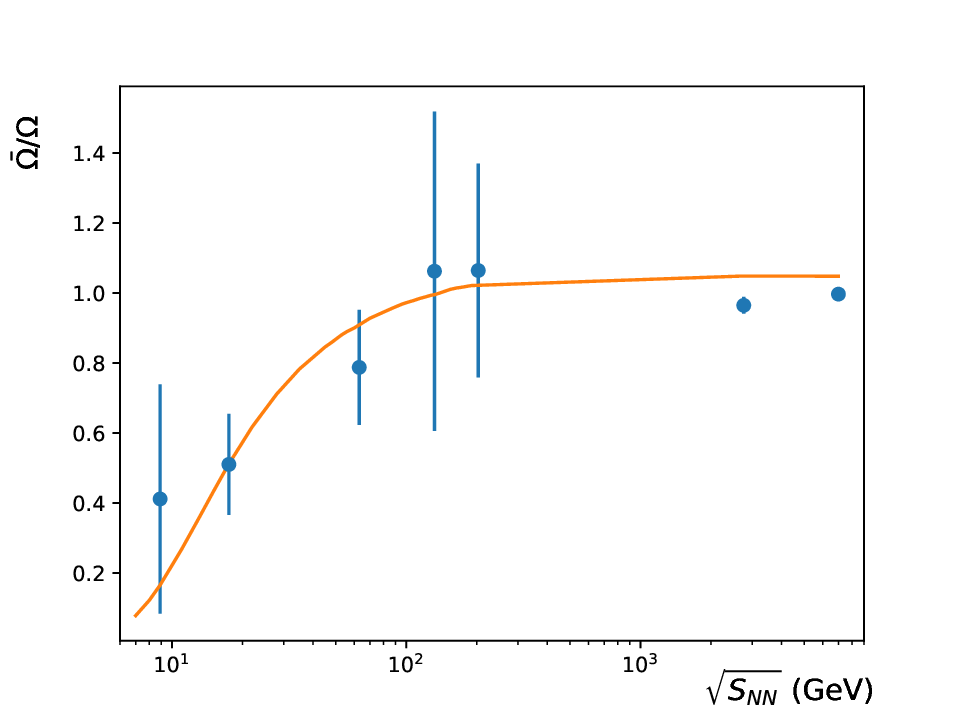}
    \caption{$\bar{\Omega}/\Omega$ dependence on  $\sqrt{s_{NN}}$. Experimental data is taken from \cite{65,67,69,70,71}.}
   \label{fig:4}
\end{figure}

\vspace{0.5mm}
\noindent  The value of the baryonic hard-core radius turns out to be 0.20 $\pm$ 0.03 fm, which is found to provide a satisfactory fit for all the  antibaryon to baryon ratios and the relative mesonic yields. This estimation of the hard-core baryonic radius is in agreement with some earlier results where it is shown that the LQCD data of different thermodynamic quantities can be described in EVHRG model with the same hard-core radius values \cite{das,bugaev}. Further the ability of HRG model at such a small radius has been found to be high enough to reproduce the lattice energy density  data \cite{tawfik}. The parameters result in a satisfactory fit for $\bar p/p$ with $\chi^{2}/dof =$ 0.48. From the Fig.~\ref{fig:1} it is evident that the ratio $\bar p/p$ ratio increases with increasing collision energy and approaches unity towards highest RHIC energies due to the increased thermal production of anti-protons compared to protons. Beyond $\sqrt{s_{NN}}$ $\approx 100$ GeV, the newly produced hadrons become dominant in the system \cite{muzinger}. At lower center of mass energies the decreasing value of ratio  $\bar p/p$ indicates an increase in \textit{net} baryon density.
\vspace{-4mm}
 With increase in $\sqrt{s_{NN}}$ all antibaryon to baryon ratios Figs.~\ref{fig:2}--\ref{fig:4} are  found to increase, suggesting a higher level of nuclear transparency in the URNNC, where strong vacuum excitation causes most of the secondary matter to form almost symmetrically between the two receding nuclei. This effect becomes stronger towards the highest RHIC and LHC energies. At lower energy collisions the participant nucleons make up a large fraction of the bulk secondary matter  because of high baryon stopping. This finding is further supported by the decreasing value of 
\noindent net-protons ($\textit{p}$ - $\bar p$) measured in the experiments with increasing collision energy. The dependence of  anti-hyperon/hyperon ratios on  energy also exhibits a mass hierarchy, wherein the saturation value of $\sim$ 1 is reached faster for more massive hyperon (anti-hyperon) species. In case of multi-strange hyperons such a phenomena underscore the preference of symmetric anti-hyperon to hyperon formation \cite{jan shabir,sam1}. Using the same set of the parameters values we get a reasonably good fit for $\bar \Lambda/\Lambda$, $\bar \Xi/\Xi$, $\bar \Omega/\Omega$ with $\chi^{2}/dof$ being 2.02, 1.5 and 0.9, respectively.

Besides the abundance of baryons and anti-baryons the kaons abundance is also an important tool to study the strangeness production and also because they are supposedly related with the physics of the postulated quark–gluon plasma and its signatures.
\begin{figure}[h]
    \centering
\includegraphics[width=0.45\textwidth, height=0.2\textheight]{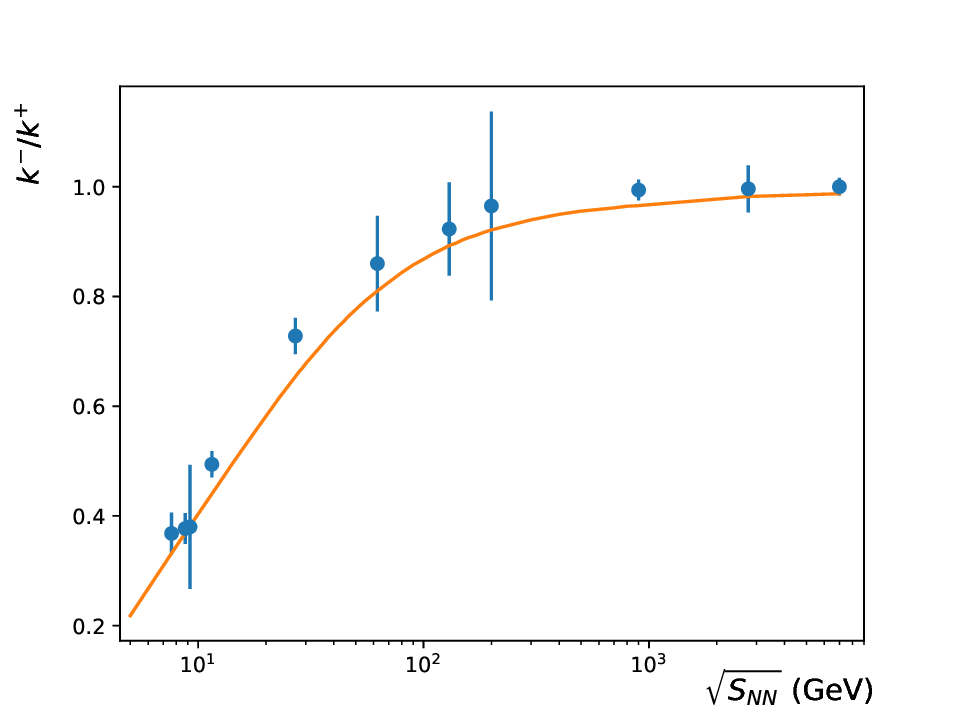}
    \caption{$k^{-}/k^{+}$ dependence on $\sqrt{s_{NN}}$. Experimental data taken from\cite{66,69,75}.}
   \label{fig:5}
\end{figure}
In Fig.~\ref{fig:5} using the same set of parameter values we have fitted $k^{-}/k^{+}$ experimental data  available over a broad range of $\sqrt{s_{NN}}$ (with $\chi^{2}/dof$ = 1.45). From the figure it is clear that $k^{-}/k^{+}$ values obtained from our theoretical model calculations increase with $\sqrt{s_{NN}}$ in accordance with the experimental values, which is similar to that of the antibaryon/baryon ratio cases. The increasing trend in the $k^{-}/k^{+}$ ratio can be again explained by the fact that $k^{-}$ besides having a strange quark (s), also contains one light anti-quark ($\bar q$). Towards the higher collision energies the production of $\bar q$ increases faster than q while the s and $\bar s$ production takes place in equal numbers due to strangeness conservation in the system \cite{cleyman}.
The explanation of the energy dependence of $k^-/k^+$ can be also be obtained from their underlying reaction mechanism in the hadronic phase. At lower and intermediate energies the system is dominated by nucleons. Consequently, the thermal reaction which predominantly produces \( k^+ \) are $N + N \rightarrow
N + X + k^+$ and $\pi + N \rightarrow X + k^+$, where N is the nucleon and X is either $\Lambda$ or $\Sigma$ hyperon. At higher energies, in addition to the above, the pair production through the reaction $\pi + \pi \rightarrow k^+ + k^-$ will also contribute significantly as the required threshold energy will be easily exceeded. Further, on the side of the baryonic sector, the process $N + \bar{N} \rightarrow k^+ + k^-$ will also start contributing as there will be an abundance of anti-nucleons in the system at the highest RHIC and LHC energies \cite{batacharya,muller, abelev}.
\begin{figure}[h]
    \centering
\includegraphics[width=0.45\textwidth, height=0.2\textheight]{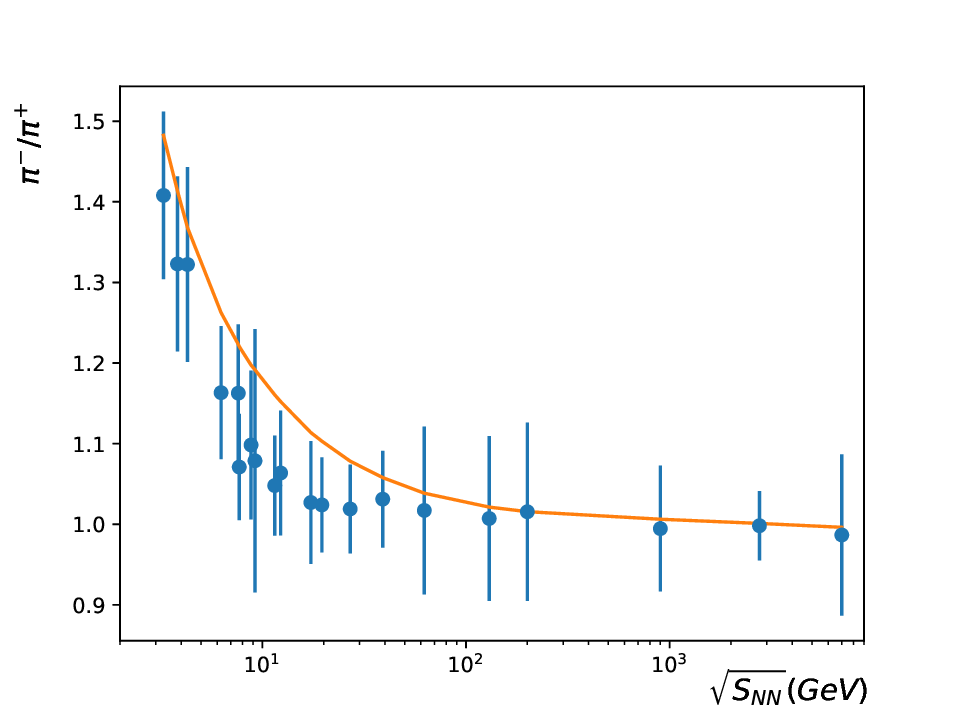}
    \caption{$\pi^{-}/\pi^{+}$ dependence on $\sqrt{s_{NN}}$. Experimental data is taken from \cite{63,69}.}
   \label{fig:6}
\end{figure}
In Fig.~\ref{fig:6} we have shown the variation of ${\pi^{-}}/{\pi^{+}}$ with $\sqrt{s_{NN}}$ to reflect the decreasing importance of the relative abundance of different iso-spin states. At higher collision energies significant contribution to the final state pion abundance comes from the secondary pions resulting from the decay of the heavier hadronic resonances \cite{h.satz}. However, at lower collision energies most of the pions produced are of direct thermal origin. The experimental data shows that this ratio is greater than unity at lower energies ($\sim$ 1.4) and approaches $\sim$ 1 as $\sqrt{s_{NN}}$ increases, which can be attributed to almost symmetric direct thermal production of $\pi^{-}$ and $\pi^{+}$ as well as through the resonance decays contributions. This could again indicate the existence of a state of a hot hadronic matter which is almost symmetric in particles and antiparticles \cite{inam}. Our theoretical values are in fairly good agreement with the experimental values for this case  (with $\chi^{2}/dof$ = 0.96).
\begin{figure}[H]
    \centering
\includegraphics[width=0.45\textwidth, height=0.2\textheight]{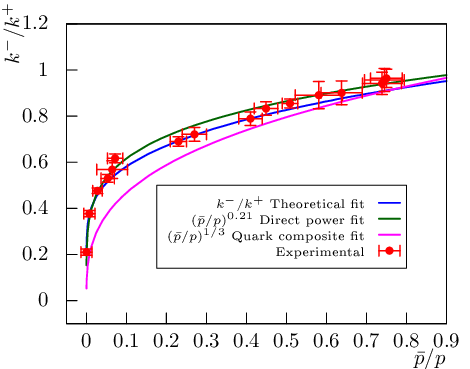}
\caption{Correlation between $k^{-}/k^{+}$ and $\bar p/p$. Experimental data is taken from \cite{69,stiles,s.das,brahms} }
\label{fig:7}
\end{figure}
The ratio {$\bar{p}/p$ can also show increasing importance of strangeness production and its evolution in the system as a function of increasing collision energy. With this aim in Fig.~\ref{fig:7} we have shown the correlation between $k^{-}/k^{+}$ and $\bar p/p$ in URNCC over a wide range of energy ($\sqrt{s_{NN}}$= 5 GeV to 2.7 TeV). This correlation indicates how kaon production is related to $net$ baryon density  and is best explained by power-law relationship as $k^{-}/k^{+} =  (\bar p/p)^{\alpha}$ \cite{song}. The blue curve shows the result of our calculations, while the pink curve is obtained by considering the light quark compositions, which provides $\alpha$ = 0.33 and fits the data poorly. For comparison, we have represented our theoretical blue curve by direct fit  using power law and
obtained  $\alpha$ = 0.23, which is close to the experimental value of $\alpha$ = 0.21  represented by the green curve. The correlation is therefore well explained by our theoretical model calculations.  
 \begin{figure}[h]
    \centering
\includegraphics[width=0.45\textwidth,height=0.2\textheight]{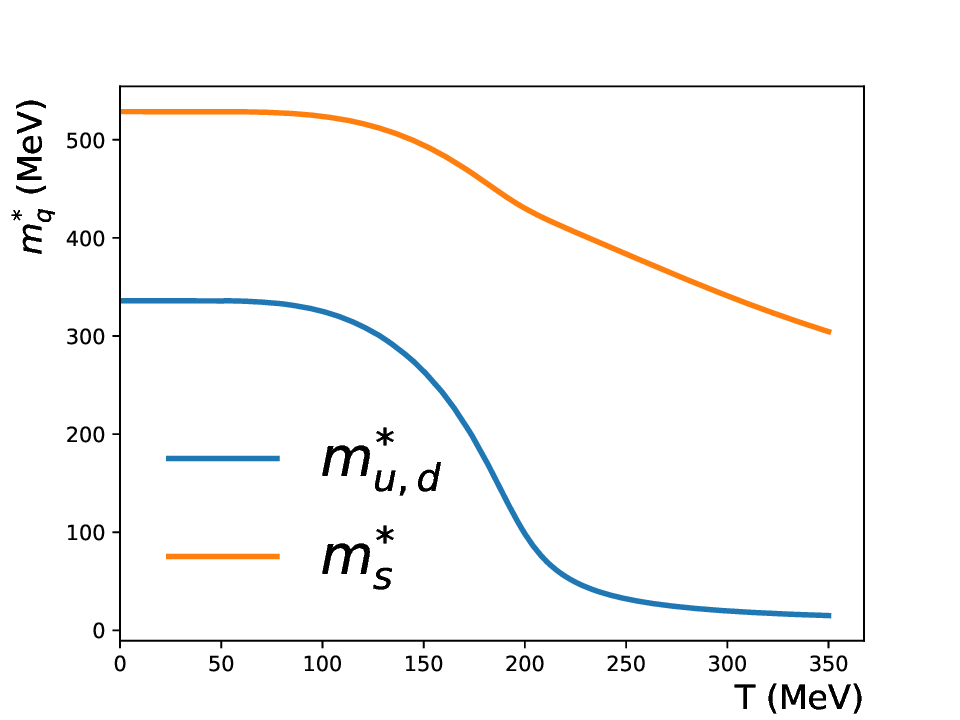}
    \caption{The graph shows the variation of constituent quark masses ($m_{q}^{*}$) as a function of Temperature (T) at $\mu=0$ under $SU(3)_{f}$ NJL  model. From the Figure it is clear that $m_{u,d}^*$ decreases
 with T and drops to current quark mass around 200 MeV, though the strange quark mass $(m_{s}^* )$ also decreases with
T, it does not attain its current quark mass.}
   \label{fig:8}
\end{figure}
\begin{figure}[h]
    \centering
\includegraphics[width=0.45\textwidth,height=0.2\textheight]{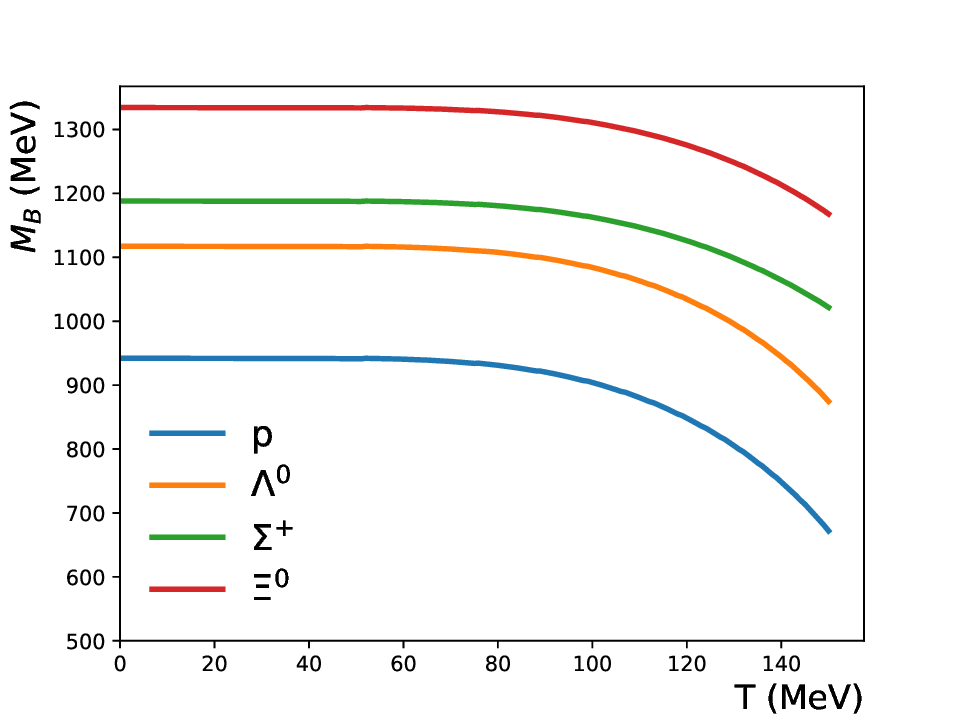}
\includegraphics[width=0.45\textwidth,height=0.2\textheight]{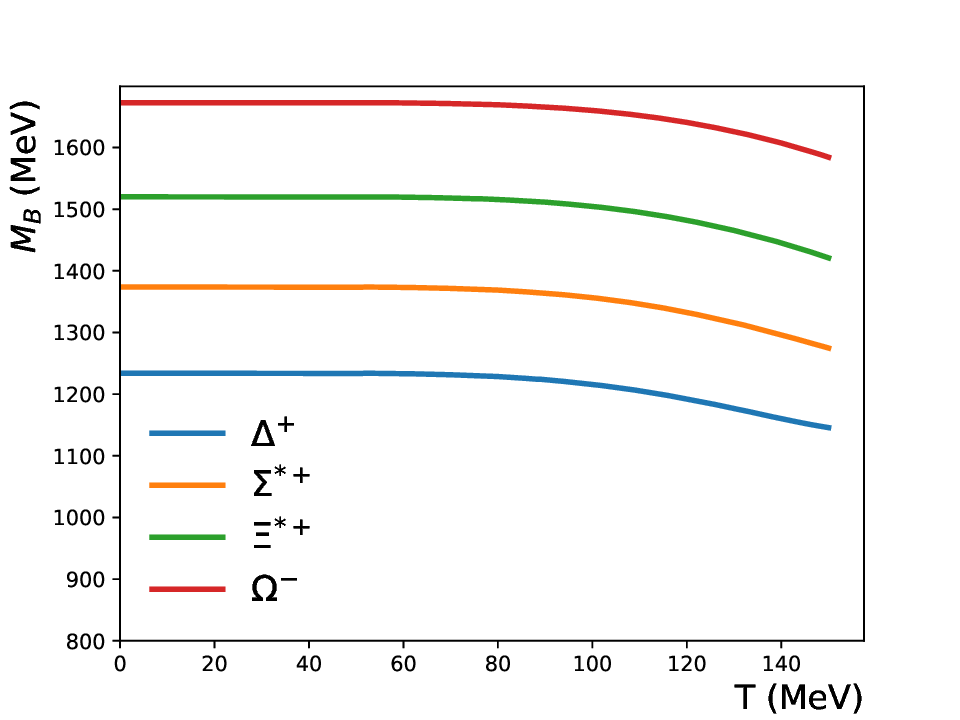}
\caption{The graphs show the variation of baryon octet and decuplet masses with temperature (T) at $\mu=0$.}
\label{fig:9}
    \end{figure}

Since in our model we have used T, $\mu$  dependent baryon masses, hence the deduced freeze-out conditions (T and $\mu_{B}$) are expected to change, compared to the cases where  hadron masses have been treated as fixed entities \cite{gast}. We have found that in the dynamical hadron mass picture, there is a lowering of all baryon masses, which leads to a relatively smaller values of chemical freeze-out temperature as compared to other models. This lowering of masses towards the highest freeze-out temperatures in our case ($\sim 145 $ MeV) is most pronounced in light mass baryons like proton (23\% lower from its vacuum mass ) followed by $\Lambda^{0}$ (18\% mass lower from its vacuum mass) while the baryon decuplet masses do not change significantly from their vacuum masses (the mass of $\Omega^{-}$ lowers only by 4\% from its vacuum mass).  This variation of baryons masses with increasing temperature is shown in Fig.~\ref{fig:9}. The relatively smaller values of chemical freeze-out temperature obtained in our case at higher collision energies ($\sim$ 145 MeV) may even be a desired effect, since higher temperatures $\sim$ 170 MeV obtained in simple thermal models that rely on vacuum hadronic masses, may correspond to  QGP phase rather than to hadron gas phase \cite{michalec}. To describe the dependence of the freeze-out parameters on the collision energy we have shown in Figs.~\ref{fig:10} and \ref{fig:11}, the dependence of T and $\mu_{B}$ on $\sqrt{s_{NN}}$. The temperature  curve increases initially and  saturates beyond $\sqrt{s_{NN}} \sim$ 40 GeV which can possibly  mean that as the collision energies increase, the system instead of increasing its temperature produces more particles. The value of $\mu_{B}$  initially decreases rapidly from a value of around 680 MeV at $\sqrt{s_{NN} }\sim 2$ GeV upto $\sqrt{s_{NN}} \sim$ 200 GeV.   Beyond which it asymptotically approaches to zero which can be attributed to nearly equal high degree of excitation in the baryon and anti-baryonic sector.
\begin{figure}[h]
    \centering
\includegraphics[width=0.45\textwidth, height=0.2\textheight]{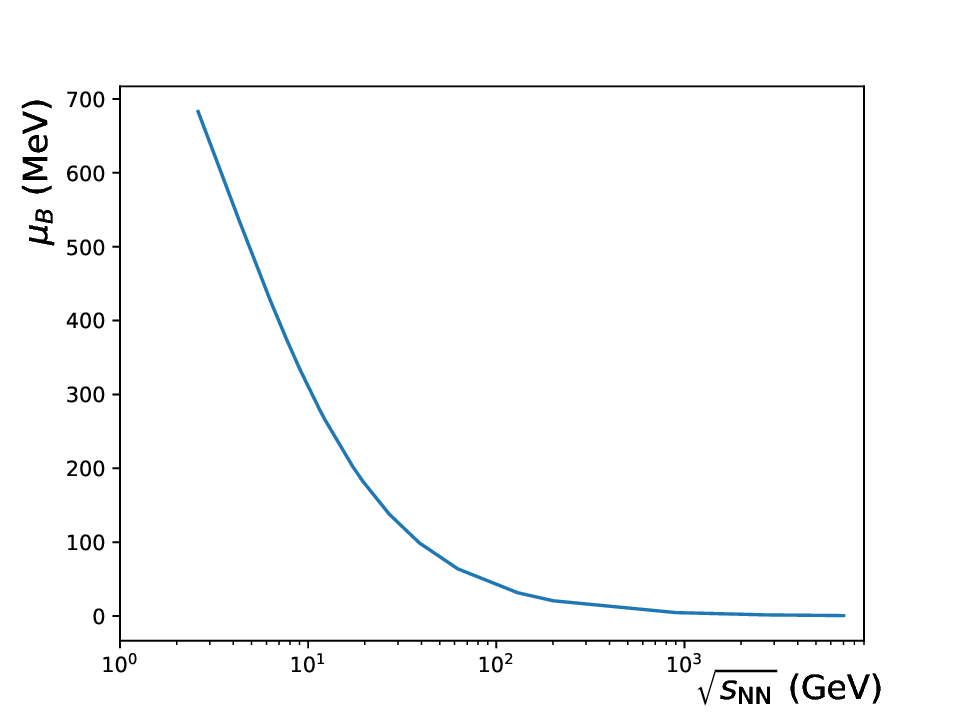}
    \caption{The graph shows the variation of $\mu_{B}$ with $\sqrt{s_{NN}}$.}
   \label{fig:10}
\end{figure}
\vspace*{2mm}
\begin{figure}[h]
    \centering
\includegraphics[width=0.45\textwidth, height=0.2\textheight]{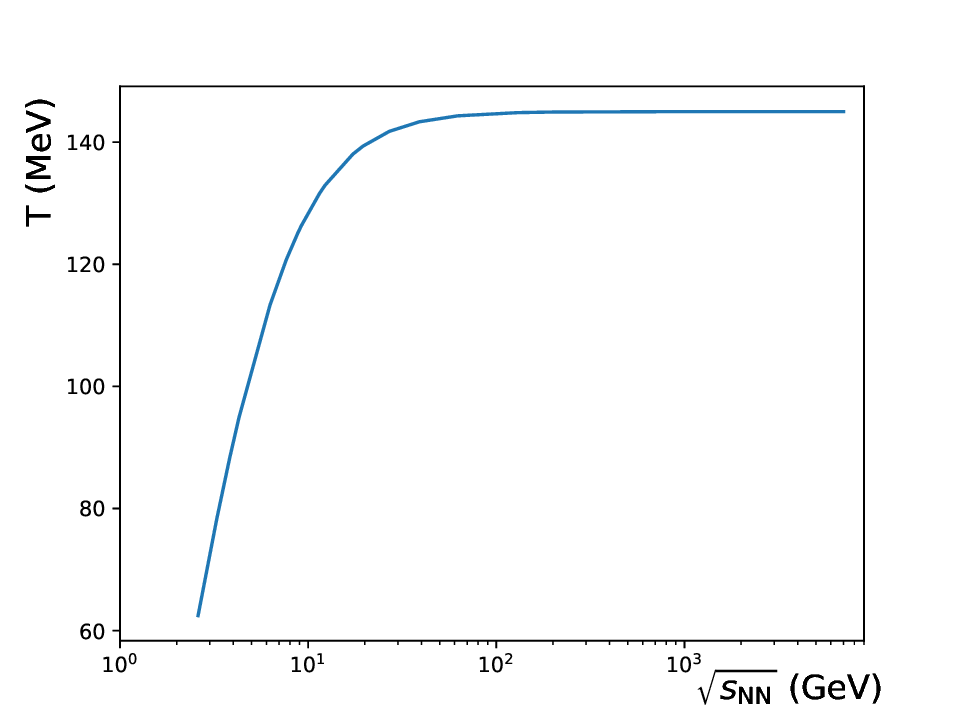}
    \caption{The graph shows the dependence of Temperature (T) on $\sqrt{s_{NN}}$.}
   \label{fig:11}
\end{figure}

 In Fig.~\ref{fig:12} we have shown a comparison of our freeze-out line with some earlier results. The figure shows to some extent a parabolic dependence of chemical freeze-out stages in the $T- \mu_{B}$ plane. While the freeze-out line in case of Cleyman et. al.\cite{cleymans freezeout paper} and Andronic et. al. \cite{sahoo} are for  point like cases, the freeze-out line of Poberezhnyuk et. al. \cite{pober} is for hard-core radius calculation. The deviation of our freeze-out line from their results can be explained on the basis of dynamical mass approach which we have incorporated in our model, while the other model calculations have used the vacuum hadronic masses. It may be also pointed that our freeze-out parameters ($T,\mu_{B}$) are dependent on the various particle 
 \begin{figure}[h]
    \centering
\includegraphics[width=0.45\textwidth, height=0.25\textheight]{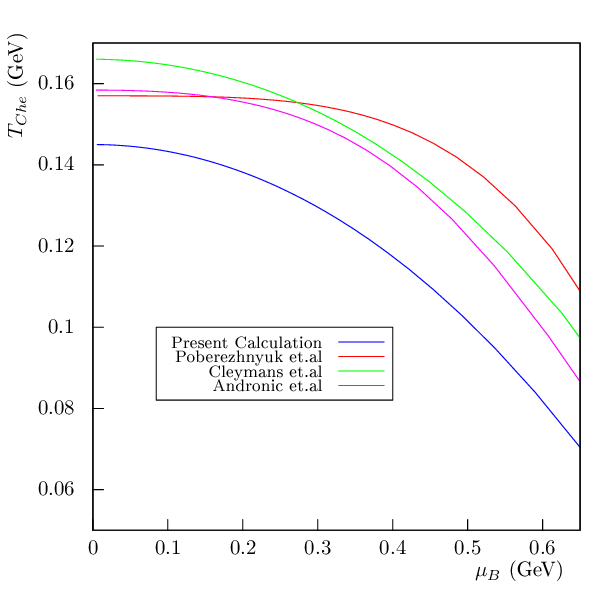}
    \caption{Chemical freeze-out line from different calculations.}
   \label{fig:12}
\end{figure}
\noindent ratios measured experimentally at different $\sqrt{s_{NN}}$. Our smaller freeze-out temperature nevertheless is in agreement with the earlier works based on effective hadron mass scaling calculations \cite{d.zsh}. Moreover Hirsch \textit{et. al.} \cite{ hirsch} in their work based on in-medium mass effect  have also found a freeze-out temperature of $\sim 136$ MeV at RHIC energy $\sqrt{s_{NN}}$= 130 GeV which is somewhat close to our model calculation value of $\sim$ 142 MeV. 
  
\section{Summary}
A number of URNNC experiments have been carried out at different energies ranging from AGS to LHC, which has made possible to  understand the equation of state of hadronic phase and the development of the EVHRG model. The present  work has focused on incorporating dynamic mass of baryons in the framework of a thermal EVHRG model by using $T,\mu$ dependent constituent quark masses  instead of simply scaling the baryon masses as had been done previously \cite{michalec,d.zsh,hirsch}. Several experimental relative hadron yields have been analyzed over a wide  range of collision energy. The quality of the fits obtained is found to be quite satisfactory, providing support for the validity of the model approach. The parametrization of temperature and baryon chemical potential showing their variation over the wide energy range of $\sqrt{s_{NN}}$ considered is established. All freeze-out parameters are extracted from the fits of the experimental data on $\bar {p}/p $ using our  theoretical results and the same are found to describe the other ratios quite satisfactorily. The correlation between $k^{-}/k^{+}$ and $\bar{p}/p$ is well described. The  extracted values of  temperature and baryon chemical potential are lower than those extracted in the other models that rely on vacuum hadron masses. The freeze-out line obtained has been compared with some previous results from which it can be concluded that the relatively lower values of temperature and baryon chemical potential in our case seem reasonable and are more likely to correspond to a hadronic phase.

\end{document}